\documentstyle[12pt]{article}
\def\be {\begin{equation}}
\def\ee {\end{equation}}
\def\ba {\begin{eqnarray}}
\def\ea {\end{eqnarray}}

\def\bi {\begin{itemize}}
\def\ei {\end{itemize}}
\begin{document}
\def\bea{\begin{eqnarray}}
\def\eea{\end{eqnarray}}
\title{\bf {Near Extremal Schwarzschild-de Sitter Black Hole Area Spectrum
from Quasi-normal Modes }}
 \author{M.R. Setare  \footnote{E-mail: rezakord@ipm.ir}
  \\{Physics Dept. Inst. for Studies in Theo. Physics and
Mathematics(IPM)}\\
{P. O. Box 19395-5531, Tehran, IRAN }}
\date{\small{}}

\maketitle
\begin{abstract}
Using the quasi-normal modes frequency of near extremal
Schwarzschild-de Sitter black holes, we obtain area and entropy
spectrum for black hole horizon. By using Boher-Sommerfeld
quantization for an adiabatic invariant  $I=\int {dE\over
\omega(E)}$, which $E$ is the energy of system and $\omega(E)$ is
vibrational frequency, we leads to an equally spaced mass
spectrum. In the other term we extend directly the Kunstatter's
approach \cite{kun}
 to determine mass and entropy spectrum of near extremal
Schwarzschild-de Sitter black holes which
 is asymptotically de Sitter rather than asymptotically flat.
 \end{abstract}
\newpage

 \section{Introduction}
 The quantization of the black hole horizon area is one of the most interesting manifestations
 of quantum gravity. Since its first prediction by
Bekenstein  in 1974 \cite{bek1}, there has been much work on this
topic \cite{bek2}-\cite{set2}. Recently, the quantization of the
black hole area has been considered \cite{hod}, \cite{dry} as a
result of the absorption of a quasi-normal mode excitation.
 Any non-dissipative systems has modes of vibrations, which forming a
complete set, and called normal modes. Each mode having a given
real frequency of oscillation and being independent of any other.
The system once disturbed continues to vibrate in one or several
of the normal modes. On the other hand, when one deals with open
dissipative system, as a black hole, instead of normal modes, one
considers quasi-normal modes for which the frequencies are no
longer pure real, showing that the system is loosing energy. In
asymptotically flat spacetimes the idea of QNMs started with the
work of Regge and Wheeler \cite{reggeW} where the stability of a
black hole was tested, and were first numerically computed by
Chandrasekhar and Detweiler several years later \cite{Chandra1}.
The quasi-normal modes bring now a lot of interest in different
contexts: in AdS/CFT correspondence
\cite{Horowitz-Habeny}-\cite{carli}, when considering
thermodynamic properties of black holes in loop quantum gravity
\cite{{dry},{kun},{mot},{hon}}, in the context of possible
connection with
critical collapse \cite{Horowitz-Habeny,BHCC,kim}.\\
 Bekenstein's idea
for quantizing a black hole is based on the fact that its horizon
area, in the nonextreme case, behaves as a classical adiabatic
invariant \cite{bek1}, \cite{bekenstescola}. In the spirit of
Ehrenfest principle, any classical adiabatic invariant corresponds
to a quantum entity with discrete spectrum, Bekenstein conjectured
that the horizon area of a non extremal quantum black hole should
have a discrete eigenvalue spectrum. Moreover, the possibility of
a connection between the quasinormal frequencies of black holes
and the quantum properties of the entropy spectrum was first
observed by Bekenstein \cite{bek4}, and further developed by Hod
\cite{hod}. In particular, Hod proposed that the real part of the
quasinormal frequencies, in the infinite damping limit, might be
related via the correspondence principle to the fundamental quanta
of mass and angular momentum. The proposed correspondence between
quasinormal frequencies and the fundamental quantum of mass
automatically leads to an equally spaced area spectrum.
Remarkably, the spacing was such as to allow a statistical
mechanical interpretation for the resulting eigenvalues for the
Bekenstein-Hawking entropy. Dreyer\cite{dry} also used the large
damping quasi-normal mode frequency to fix the value of the
Immirzi parameter, $\gamma$, in loop quantum gravity. He found
that loop quantum gravity gives a correct prediction for the
Bekenstein-Hawking entropy if gauge group
should be SO(3), and not SU(2).\\
In this letter our aim is to obtain the area and entropy spectrum
of near extremal Schwarzschild- de Sitter black holes in four
dimensional spacetime. According a recent study
\cite{Cardoso}(based on identifying the relevant scattering
potential with that of the Poscl-Teller model \cite{{PT},{FM}})
the quasinormal spectrum of SdS space depends strongly on the
orbital angular momentum of the perturbation field. Moreover, in
the near extremal SdS case, the real part of the frequency, goes
almost linearly with orbital angular momentum. In a more  recent
paper \cite{ab1}, the quasinormal mode spectrum was calculated for
 SdS space by way of the monodromy method
\cite{motl}. When this form of the spectrum is then subjected to
the near extremal  limit, as was done explicitly in \cite{ab1},
there  is absolutely no orbital angular momentum dependence in
evidence.

\section{Near Extremal Schwarzschild- de Sitter Black Hole}
The metric of the Schwarzschild-de Sitter spacetime is as
following \be ds^{2}=-f(r)dt^{2}+f^{-1}(r)dr^{2}+r^{2}d\Omega^{2},
\label{metr} \ee where \be
f(r)=1-\frac{2M}{r}-\frac{r^{2}}{a^{2}}, \label{metel} \ee where
$M$ is the mass of black hole, and $a$ denoting the de Sitter
curvature radius, which related to the cosmological constant by
$a^{2}=\frac{3}{\Lambda}$. The spacetime possesses two horizons,
the black hole horizon is at $r=r_{b}$, and the cosmological
horizon is at $r=r_{c}$, with $r_b< r_c$. The third zero of
$f(r)$ locates at $r_0=-(r_b+r_c)$. It is useful to express $M$
and $a^{2}$ as function of $r_b$, $r_c$ \be
a^2=r_{b}^{2}+r_br_c+r_{c}^{2} \label{aeq} \ee and \be
2Ma^2=r_br_c(r_b+r_c) \label {meq} \ee The surface gravity $k_b$
associated with the black hole horizon, as defined by the
relation $k_b=\frac{1}{2}\frac{df}{dr}|_{r=r_b}$. Easily one can
show \be k_b=\frac{(r_c-r_b)(r_b-r_0)}{2a^{2}r_b}.
 \label{surgr}\ee Let us now specialize to the near extremal SdS
 black hole, which is defined as the spacetime for which the
 cosmological horizon is very close to the black hole horizon. In
 this case we have
 \be r_0\sim -2r_b, \hspace{1cm} a^2\sim 3r_{b}^{2} , \hspace{1cm}k_b\sim
 \frac{r_c-r_b}{2r_{b}^{2}}. \label{apro} \ee
 The analytical quasinormal mode spectrum for the near extremal SdS
black hole has been derived by Cardoso and Lemos \cite{car} is as
following \be
\omega=k_b[-(n+1/2)i+\sqrt{\frac{v_0}{k_{b}^{2}}-1/4}+o[\Delta]],
\hspace{1cm}n=0,1,2,... \label{quasi} \ee where \be
v_0=k_{b}^{2}l(l+1), \label{scaler} \ee for scalar and
electromagnetic perturbations, and \be v_0=k_{b}^{2}(l+2)(l-1),
\label{gravi} \ee for gravitational perturbations, $l$ is the
angular quantum number. In Eq.(\ref{quasi}) $\Delta$ is squeezing
parameter \cite{med} which is as following \be
\Delta=\frac{r_c-r_b}{r_b}\ll 1. \label{deleq} \ee  Let
$\omega=\omega_{R}-i\omega_{I}$, then $\tau=\omega_{I}^{-1}$ is
the effective relaxation time for the black hole to return to a
quiescent state. Hence, the relaxation time $\tau$ is arbitrary
small as $n\rightarrow \infty$. We assume that this classical
frequency plays an important role in the dynamics of the black
hole and is relevant to its quantum properties \cite{{hod},{dry}}.
In particular, we consider $\omega_{R}$ the real part of
$\omega$, to be a fundamental vibrational frequency for a black
hole of energy $E$. Here if we adopt the mass definition due to
the Abbott and Deser \cite{de} we have $E=M$. Given a system with
energy $E$ and vibrational frequency $\omega$ one can show that
the quantity \be I=\int {dE\over \omega(E)}, \label{bohr} \ee is
an adiabatic invariant \cite{kun}, which via Bohr-Sommerfeld
quantization has an equally spaced spectrum in the semi-classical
(large $n$) limit: \be I \approx n\hbar. \label{smi} \ee Now by
taking $\omega_{R}$ in this context, we have \be I=\int
\frac{dE}{\omega_{R}}=\int \frac{dM
}{k_b\sqrt{\frac{v_0}{k_{b}^{2}}-1/4}} =
\frac{M}{k_b\sqrt{\frac{v_0}{k_{b}^{2}}-1/4}}+c, \label{adin} \ee
where $c$ is a constant. Boher-Sommerfeld quantization then
implies that the mass spectrum is equally spaced, \be M=n\hbar
k_b\sqrt{\frac{v_0}{k_{b}^{2}}-1/4}. \label{masspec}\ee The use
of Eqs.(\ref{meq},\ref{aeq}) leads us to \be \delta M=\frac{r_b
\Delta r \delta r_{b}}{2a^{2}}=\hbar
k_b\sqrt{\frac{v_0}{k_{b}^{2}}-1/4}. \label{delmas} \ee In the
other hand, the black hole horizon area is given by \be A_b=4\pi
r_{b}^{2}, \label{area} \ee by the variation of the black hole
horizon and use of Eq.(\ref{delmas}) we have \be \delta A_b=8 \pi
r_b \delta r_b=8\pi \frac{2a^2\hbar
k_b\sqrt{\frac{v_0}{k_{b}^{2}}-1/4}}{\Delta r}. \label{delare}
\ee Now by using Eq.(\ref{area}), one can obtain \be \delta
A_b=24 \pi \hbar\sqrt{\frac{v_0}{k_{b}^{2}}-1/4}. \label{delare1}
\ee Then we can obtain the quantization of a near extremal SdS
black hole area as \be A_b=24 \pi n
\hbar\sqrt{\frac{v_0}{k_{b}^{2}}-1/4}. \label{areasp} \ee
 Using the definition of the
Bekenstein-Hawking entropy we have \be S=\frac{A_{b}}{4\hbar}=6\pi
n \sqrt{\frac{v_0}{k_{b}^{2}}-1/4}. \label{entro} \ee
  \section{Conclusion}
Bekenestein's  idea for quantizing a black hole is based on the
fact
  that its horizon area, in the nonextremal case,
behaves as a classical adiabatic invariant. It is interesting to
investigate how near extremal black holes would be quantized.
Discrete spectra arise in quantum mechanics in the presence of a
periodicity  in the classical system Which in turn leads to the
existence of an adiabatic invariant  or action variable.
Boher-Somerfeld quantization implies that this adiabatic invariant
has an equally spaced spectrum in the semi-classical limit.
Kunstatter showed that this approach works for the
 Schwarzschild black holes in any dimension, giving asymptotically equally
 spaced areas, here we showed generalization to near extremal SdS black
 holes is also successful. In
this letter we have considered the near extremal SdS black hole in
four dimensional spacetime, using the results  for highly damped
quasi-normal modes, we obtained the area and entropy spectrum of
event horizon. We have found that the quantum area is $\Delta
A=24 \pi \hbar\sqrt{\frac{v_0}{k_{b}^{2}}-1/4}$. As one can see
for example in \cite{{kun},{bir},{and}}, $\Delta A=4\hbar \ln3$
for schwarzschild black hole. Therefore in contrast with claim of
\cite{and}$, \Delta A $ is not universal for all black holes.
Also Abdalla et al \cite {ab} have shown that the results for
spacing of the area spectrum for near extreme Kerr  black holes
differ from that for schwarzschild, as well as for non-extreme
Kerr black holes. Such a difference for problem under
consideration in this letter also in \cite{ab} as the authors have
been mentioned may be justified due to the quite different nature
of the asymptotic quasi-normal mode spectrum of the near extreme
black hole.

\end{document}